\documentclass[a4paper,11pt]{article}

\usepackage{amsfonts,amsmath, color}
\usepackage{bm, bbm}
\usepackage{graphicx}
\usepackage{mathtools}

\newcommand{\be}{\begin{equation}}
\newcommand{\bal}{\begin{align}}
\newcommand{\eal}{\end{align}}
\newcommand{\ee}{\end{equation}}
\newcommand{\bea}{\begin{eqnarray}}
\newcommand{\eea}{\end{eqnarray}}
\newcommand{\bit}{\begin{itemize}}
\newcommand{\eit}{\end{itemize}}
\newcommand{\la}{\langle}
\newcommand{\ra}{\rangle}

\usepackage{color}

\usepackage{jheppub} 

\usepackage[T1]{fontenc} 

\title{Horizon temperature without space-time\footnote{Talk presented at the Corfu Summer Institute 2019 ``School and Workshops on Elementary Particle Physics and Gravity"
		31 August - 25 September 2019
		Corfu, Greece.}}

\author[a,b]{Michele Arzano}

\affiliation[a]{Dipartimento di Fisica ``E. Pancini",\\ Universit\`a di Napoli Federico II, I-80125 Napoli, Italy}
\affiliation[b]{INFN, Sezione di Napoli,\\ Complesso Universitario di Monte S. Angelo, Via Cintia Edificio 6,\\ 80126 Napoli, Italy}

\emailAdd{michele.arzano@na.infn.it}

\abstract{It is shown how the characteristic thermal effects that observers experience in space-times possessing an event horizon can manifest already in a simple quantum system with affine symmetry living on the real line. The derivation presented is essentially group theoretic in nature: a thermal state emerges naturally when comparing different representations of the group of affine transformations of the real line. The freedom in the choice of different notions of translation generators is the key to the Unruh effect ``on a line" we describe.}

\begin{document} 
\maketitle
\flushbottom

\section{Introduction}

Does the vacuum state of a quantum field look ``empty", i.e. devoid of particles, to all observers? As noticed in \cite{Davies:1984rk} this simple yet fundamental question did not receive much attention until the 1970s when it was realized that different vacuum states can be constructed for each choice of time-like Killing vector generating time evolution  \cite{Ashtekar:1975zn} . As it happens this rather abstract observation is directly related to one of the most significant insights semiclassical gravity provides on quantum gravity. This goes back to Bekenstein, who, in the early 1970s, based on the enigmatic analogy between the laws of black hole mechanics and those of thermodynamics \cite{Bekenstein:1973ur} proposed to associate an entropy to black holes proportional to their area divided by the Planck length ($L_p \sim 10^{-35}$ m) squared. This suggestive conjecture was soon spectacularly confirmed by Hawking \cite{Hawking:1974sw} who showed that static observers far away from a black hole detect thermal radiation at temperature $T_{H} = \frac{1}{8\pi GM}$ where $M$ is the black hole mass and $G$ is Newton's constant. At the basis of this phenomenon lies the fact that the vacuum state for free falling observers {\it does not} coincide with the vacuum state for static observers which perceive the former as a thermal state at the Hawking temperature $T_H$.

It turns out that such thermal behaviour is not exclusive to black hole horizons but is much more general and it manifests in {\it any} space-time admitting a causal horizon \cite{Hollands:2014eia}. In particular even in ordinary flat Minkowski space-time uniformly accelerated observer have causal access only to a portion of space-time, the Rindler wedge, bounded by a horizon, and  perceive what is the vacuum state for inertial observers as a thermal state at a temperature proportional to the magnitude of their acceleration. This effect was discovered by Unruh \cite{Unruh:1976db} soon after Hawking's work and, like the effect discovered by Hawking, has its roots in the different choices of generators of time translations associated to two classes of observers: for inertial observers the proper time flow along their worldlines is generated by the time-like element of the abelian subalgebra of the Poincar\'e algebra, while time evolution for uniformly accelerated observers is determined by the generator of boosts along the direction of acceleration belonging to the Lorentz subalgebra of the Poincar\'e algebra.

In this contribution we will focus on the group theoretic aspects underlying these ``horizon temperature" phenomena and show how they can manifest even in the absence of a space-time equipped with a metric and causal horizons \cite{Arzano:2018oby}. We will consider a quantum system living on the real line and assume that its symmetries are given by the group of affine transformations: translations and dilation. Both generators of these transformations can be considered as describing time evolution of the system and a Unruh-type effect can take place when describing the vacuum state associated to translations in terms of excitations related to the dilation generator. This very minimal setting can be actually seen as a toy version of a quantum system (field theory) invariant under the Poincar\'e group. Indeed the group of affine transformations, like the Poincar\'e group, is a semidirect product group with dilations playing the role of Lorentz transformations acting on the translation sector. As in standard quantum field theory, we will consider irreducible representations of the affine group as elementary excitations or ``particles". Thus our starting point will be a brief description of the affine group and its irreducible representations.

\section{The $ax+b$ group and its representations}

The group of affine transformation of the real line, also known in the mathematical literature as the ``$ax+b$'' group, consists of combinations of translations and dilations 
\be\label{ax+b}
x\rightarrow ax+b\,,\,\,\,\,\, a\in \mathbb{R}^+\,,\,\, b\in \mathbb{R}\,.
\ee
Setting $a=1$ we obtain the subgroup of translations of the real line; the action of this subgroup is transitive since for any given $x$ all elements of $\mathbb{R}$ can be reached through a translation by a given $b$. The subgroup of dilations defined by $b=0$ is transitive only on $\mathbb{R}^+$ (or $\mathbb{R}^-$). This subgroup can be regarded as a group of translations of the real positive (negative) semi-line. Therefore the $ax+b$ group can be seen as a group comprised of two types of translations, one on the real line and the other on half of it (positive or negative). Denoting a generic group element as $g=(a,b)$ the group multiplication law for two elements $g_1=(a_1,b_1)$ and $g_2=(a_2,b_2)$ is given by
\be
g_1 g_2 = (a_1 a_2, a_1 b_2+b_1)\,.
\ee
The unitary irreducible representations of the $ax+b$ group are well known (see e.g. \cite{VileK, Moses:1974ez}). There are just two such representations. One is realized on the Hilbert space $\mathcal{H}^+_{\omega}$ of functions on the positive real $\omega \in \mathbb{R}$ line whose elements we denote with $|\omega\ra_+$. The actions of translations $T(b) = (1, b)$ and dilations $D(a) = (a, 0)$ on such Hilbert space are given by \cite{VileK}
\be
T(b) |\omega\ra_+ = e^{- i b \omega}\,  |\omega\ra_+\,,\qquad D(a) |\omega\ra_+ =  |a\, \omega\ra_+\,.
\ee
and these can be rewritten in terms of the generators $P$ and $R$ as 
\bea\label{tidk}
T(\alpha) |\omega\ra_+  &= e^{-i \alpha P}\,  |\omega\ra_+ = e^{-i \alpha \omega}\,  |\omega\ra_+ \\
D(\lambda) |\omega\ra_+ & = e^{-i \lambda R}\,  |\omega\ra_+ =  |e^{\lambda}\, \omega\ra_+ \label{dk} \,.
\eea
Differentiating with respect to the transformation parameters $\alpha$ and $\lambda$ and setting them to zero we obtain the action of the generators
\be\label{pk+}
P\, |\omega\ra_+ = \omega\, |\omega\ra_+
\ee
and
\be\label{rk+}
R\,  |\omega\ra_+ = i\omega\frac{d}{d\omega}\, |\omega\ra_+\,,
\ee
with $\omega\in \mathbb{R}^+$. These actions lead to the commutator
\begin{equation}\label{axbalg}
  [R,P]=iP
\end{equation}
among the generators which defines the Lie algebra of the $ax+b$ group, the {\it simplest} non-abelian Lie algebra. 
The action of translations and dilations on functions $\psi(\omega) = {}_+\la \omega| \psi\ra$ 
belonging to the $P$-particle Hilbert space is explicitely given by
\be\label{groupfunp}
T(\alpha)\,  \psi(\omega) = e^{i \alpha \omega}\,  \psi(\omega)\,,\,\,\,\,\,  D(\lambda)\,  \psi(\omega) = \psi(e^{-\lambda} \omega)\,,
\ee
and the invariant inner product on such space is defined using the invariant momentum space measure $\frac{d\omega}{\omega}$
\be\label{innprod+}
\langle \psi | \psi' \rangle = \int^{\infty}_0 \frac{d\omega}{\omega}\, \bar{\psi}(\omega) \psi'(\omega) =  \int^{\infty}_0 \frac{d\omega}{\omega}\, \langle\psi| \omega\rangle_+ {}_+\langle \omega| \psi'\rangle\,.
\ee
The other unitary irreducible representation is realized on the Hilbert space $\mathcal{H}^-_{\omega}$ of functions on the negative real line, whose basis kets we denote  $|\omega\ra_-$ on which the generators $P$ and $R$ act as in \eqref{pk+}, \eqref{rk+}, and with inner product given by
\be\label{innprod-}
\langle \psi | \psi' \rangle = \int^{0}_{-\infty} \frac{d\omega}{\omega}\, \bar{\psi}(\omega) \psi'(\omega) =  \int^{0}_{-\infty} \frac{d\omega}{\omega}\, \langle\psi| \omega\rangle_- {}_-\langle \omega| \psi'\rangle\,.
\ee
As mentioned in the Introduction the transformations \eqref{tidk} and \eqref{dk} can be seen as the one-dimensional counterparts of infinitesimal Poincar\'e transformations with translations generated by $T$ and the dilations $D$ playing the role of boosts generators. Following this analogy we will interpret the states $|\omega\ra_+$ belonging to the irreducible representation of the group with positive eigenvalues for the generator $P$, as ``$P$-particle states". 

\section{Affine fields}

We will think of our physical system as a $0+1$-dimensional quantum field theory, i.e. quantum mechanics, in which the field depends only on a time variable $t\in\mathbb{R}$. A generic function of the time variable can be expanded in terms of the irreducible representations above as
\cite{Moses:1974ez}
\be\label{fieldx2}
\psi(t)=\la t|\psi\ra = \frac{1}{\sqrt{2 \pi}} \int^{\infty}_{0}\, \frac{d\omega}{\omega}\, e^{i\omega t}\,  {}_+\langle \omega| \psi\rangle + \frac{1}{\sqrt{2 \pi}} \int_{-\infty}^{0}\, \frac{d\omega}{|\omega|}\, e^{i\omega t}\, {}_-\langle \omega| \psi\rangle
\ee
from which, using \eqref{groupfunp}, one obtains
\be\label{1geoax}
T(\alpha)\, \psi(t) =  \psi(t + \alpha) \,,\quad  D(\lambda)\,  \psi(t) = \psi(e^{\lambda} t)\,,
\ee
so that
\begin{equation}\label{pronx}
  P= -i\frac{d}{dt}\,,\quad R= - it\frac{d}{dt}\,,
\end{equation}
providing a time representation of the algebra \eqref{axbalg}. Notice how from \eqref{1geoax} it is evident that, while ordinary time evolution generated by $P$ spans the whole real line, adopting $R$ as a generator of time translations restricts the range of time evolution to a half line. Let us note in passing that, if $R$ is to be interpreted as a generator of time evolution, it should, like $P$, carry dimensions of inverse time, contrary to what eq.\ (\ref{pronx}) suggests. For the moment being, in order to make more readable the formulae below, we will keep this generator dimensionless, restoring its physical dimension later on in the discussion.

From \eqref{fieldx2} we see that the time representation wave-functions associated to $P$-particle states are given by positive and negative frequency plane waves
\be
\la t| \omega\ra_{\pm} =  \frac{1}{\sqrt{2 \pi}}\, e^{i\omega t}\,,\quad \omega\in \mathbb{R}^{\pm}\,,
\ee
which are the counterpart of the usual positive and negative-energy plane waves representing one-particle states and their antiparticle counterparts in quantum field theory. The analogy can be made more explicit if we change integration variable in the second term of \eqref{fieldx2} and write
\be\label{fieldx2a}
\la t|\psi\ra = \frac{1}{\sqrt{2 \pi}} \int^{\infty}_{0}\, \frac{d\omega}{\omega}\left( e^{i\omega t}\,  {}_+\langle \omega| \psi\rangle + e^{-i\omega t}\, {}_-\langle -\omega| \psi\rangle\right)\,.
\ee
For a real field $\psi(t) \equiv \psi*(t)$ and thus we must require that ${}_-\langle -\omega| \psi\rangle = ({}_+\langle \omega| \psi\rangle)^*$, and thus denoting ${}_+\langle \omega| \psi\rangle\equiv a(\omega)$ we can write
\be\label{psikx}
\psi(t)\equiv\la t|\psi\ra = \frac{1}{\sqrt{2 \pi}} \int^{\infty}_{0}\, \frac{d\omega}{\omega}\left( e^{i\omega t}\, a(\omega) + e^{-i\omega t}\, a^*(\omega) \right)\,,
\ee
an expression formally analogous to the well known expression for a real classical (on-shell) scalar field on Minkowski space-time. It must be noted that the expansion \eqref{fieldx2} carries a representation of the $ax+b$ group which is {\it reducible}. Indeed the positive and negative energy contribution are invariant subspaces under the action of dilations. Restricting to positive or negative energies one obtains an irreducible representation on the whole time line $t\in \mathbb{R}$. In particular pure positive frequency functions are elements of $P$-particle states in a time representation
\be
\la t|\psi\ra_+ = \frac{1}{\sqrt{2 \pi}} \int^{\infty}_{0}\, \frac{d\omega}{\omega}  e^{i\omega t}\,  {}_+\langle \omega| \psi\rangle\,.
\ee
Starting from such one-particle Hilbert space one can proceed with the usual Fock space construction \cite{Geroch:2013i}. The P-vacuum state is then defined by the requirement
\be
a(\omega) |0\ra_P = 0\,,
\ee
where $a(\omega)$ is the annihilation operator obtained quantizing the coefficient ${}_+\langle \omega| \psi\rangle\equiv a(\omega)$.

\section{Mellin transform and $R$-eigenstates}

We now introduce a set of states which diagonalize the dilation generator $R$. In analogy with the states $|\omega\ra_{\pm}$, which carry a charge (frequency) associated to the translational symmetry generated by $P$, these states will carry a charge associated with the {\it multiplicative} translational symmetry generated by $R$ on the positive half line $\mathbb{R}^+$. From representation theory \cite{VileK} we known that functions which diagonalize $R$ can be obtained via the Mellin transform of functions defined on the positive half line, in our case functions on the Hilbert space spanned by $|\omega\ra_+$ 
\be\label{omegamellpsi}
\la \Omega | \psi\ra =  \frac{1}{\sqrt{2 \pi}} \int^{\infty}_0\, \frac{d\omega}{\omega}\, \omega^{-i \Omega}\, {}_+\la \omega| \psi\ra\,,\qquad \Omega\in\mathbb{R}\,,
\ee
so that in bra-ket notation we have
\be\label{omk+}
\langle \Omega  | \omega \ra_+ = \frac{1}{\sqrt{2 \pi}}\, \omega^{-i \Omega}\,.
\ee
A key point to notice is that \eqref{omegamellpsi} can be used to construct annihilation operators $b(\Omega) = \la \Omega | \psi\ra $ which share the same vacuum as the $a(\omega)$ operators
\be
b(\Omega) |0\ra_P =  \frac{1}{\sqrt{2 \pi}} \int^{\infty}_0\, \frac{d\omega}{\omega}\, \omega^{-i \Omega}\, a(\omega) |0\ra_P = 0\,. 
\ee
The relation \eqref{omegamellpsi} also allows us to express an $|\Omega\ra $ state as a superposition of $P$-particle sates 
\be\label{omegamell}
|\Omega\ra = \int^{\infty}_0\, \frac{d\omega}{\omega}\, \omega^{i \Omega}\, | \omega \rangle_+\,,
\ee
from which, using \eqref{rk+}, we can write the following actions of the $ax+b$ generators
\begin{equation}\label{pronomega}
  P\, |\Omega\rangle=|\Omega+ i \rangle\,,\quad R\, |\Omega\rangle= \Omega\, |\Omega\rangle\,,
\end{equation}
showing that the $|\Omega\rangle$ states diagonalize the operator $R$, as expected. Using the inverse Mellin transform and ignoring convergence issues we can use \eqref{omegamell} to write a $P$-particle state as a superposition of  $|\Omega\rangle$ states
\be\label{invmetransk}
 |\omega\rangle_{+} = \int^{\infty}_{-\infty}\, d\Omega\, \langle\Omega |\omega \rangle_+\, |\Omega\rangle = \frac{1}{\sqrt{2 \pi}} \int^{\infty}_{-\infty}\, d\Omega\, \omega^{-i\Omega}\, |\Omega\rangle \,,\\\\\\\\\\\ \ \ \ \ \ \ \omega\in\mathbb{R}^{+}\,.
\ee
Notice that making use of the distributional identities
\be
 \int^{\infty}_{-\infty}\, d\Omega\, \omega^{i\Omega - 1}\, (\omega')^{-i \Omega} = 2 \pi\, \delta(\omega-\omega')\,,
\ee
and
\be
 \int^{\infty}_0\, \frac{d \omega}{\omega}\, \omega^{-i(\Omega - \Omega')}\, = 2 \pi\, \delta(\Omega - \Omega')\,,
\ee
it can be easily checked that the {\it covariant} inner product for $|\omega\ra_+$ states
\be
{}_+\la \omega| \omega'\ra_+ = \omega\, \delta(\omega-\omega')
\ee
is compatible with standard inner product
\be\label{omegainner}
\la \Omega|\Omega'\ra = \delta(\Omega-\Omega')
\ee
for $|\Omega\ra$ states. The Mellin transform \eqref{omegamellpsi} is an {\it isometry} between the Hilbert space spanned by the kets $| \omega\ra_+,\,\, \omega\in \mathbb{R}^+ $ with inner product \eqref{innprod+} and the Hilbert space spanned by kets $|\Omega\ra ,\,\, \Omega\in \mathbb{R}$ with inner product
\be\label{innprodome}
\langle \psi | \psi' \rangle =   \int^{\infty}_{-\infty}\, d\Omega\, \langle\psi| \Omega\rangle \langle \Omega| \psi'\rangle\,.
\ee
We can now re-write the field $\psi(t)=\la t|\psi\ra$ living on the whole real line in terms of $R$-eigenmodes $\la\Omega|\psi\rangle$. We start by writing equation \eqref{fieldx2a} as
\be
\la t|\psi\ra = \frac{1}{\sqrt{2 \pi}} \int^{\infty}_{0}\, \frac{d\omega}{\omega}\, \left(e^{i\omega t}\,  {}_+\langle \omega| \psi\rangle + e^{-i\omega t}\, {}_+\langle \omega| \psi\rangle^*\right)\,,
\ee
and substitute in it the expressions for  ${}_{+}\la \omega|\psi\rangle$ which can be obtained from \eqref{invmetransk}
\be\label{fieldx3}
\la t|\psi\ra = \frac{1}{2 \pi} \int^{\infty}_{0}\, \frac{d\omega}{\omega}\,\int^{\infty}_{-\infty}\, d\Omega\,  \left( e^{i\omega t}\, \omega^{i\Omega}\, \la \Omega|\psi\rangle +  e^{- i\omega t}\,\omega^{- i\Omega}\, \la \Omega|\psi\rangle^* \right)\,.
\ee
With the help of the identity
\be
\int_0^\infty \frac{d\omega}{\omega}\, \omega^z \, e^{\pm i\omega t} = \Gamma(z)\, t^{-z} e^{\pm i\pi z/2}\,,
\ee
we can write 
\be\label{fieldx3a}
\la t|\psi\ra = \frac{1}{2 \pi} \int^{\infty}_{-\infty}\, d\Omega\,  \left( t^{-i\Omega}\, e^{-\frac{\pi \Omega}{2}} \Gamma(i\Omega) \la \Omega|\psi\rangle +  t^{i\Omega}\, e^{-\frac{\pi \Omega}{2}} \Gamma(-i\Omega)\, \la \Omega|\psi\rangle^* \right)\,,
\ee
which can be expanded in an expression involving only {\it positive} $\Omega$
\begin{align}\label{fieldx4a}
\psi(t)=\la t|\psi\ra & =\frac{1}{2 \pi} \int^{\infty}_{0}\, {d\Omega}\, t^{-i\Omega}\Gamma(i\Omega)\left( e^{-\pi\Omega/2} \la \Omega|\psi\rangle+ e^{\pi\Omega/2} \la -\Omega|\psi\rangle^* \right) \nonumber\\
&+\frac{1}{2 \pi} \int^{\infty}_{0}\, {d\Omega}\, t^{i\Omega}\Gamma(-i\Omega)\left( e^{-\pi\Omega/2} \la \Omega|\psi\rangle^*+ e^{\pi\Omega/2} \la -\Omega|\psi\rangle \right)\,.
\end{align}
In analogy with the identification made above ${}_+\langle \omega| \psi\rangle\equiv a(\omega)$, we now denote $\la \pm \Omega|\psi\rangle = b_{\pm}(\Omega)$ so that 
\begin{align}\label{fieldx5a}
\psi(t)=\la t|\psi\ra & =\frac{1}{2 \pi} \int^{\infty}_{0}\, {d\Omega}\, t^{-i\Omega}\, \Gamma(i\Omega)\left( e^{-\pi\Omega/2}\, b_+(\Omega) + e^{\pi\Omega/2}\, b^*_-(\Omega) \right) \nonumber\\
&+\frac{1}{2 \pi} \int^{\infty}_{0}\, {d\Omega}\, t^{i\Omega}\, \Gamma(-i\Omega)\left( e^{-\pi\Omega/2}\, b^{*}_+(\Omega)+ e^{\pi\Omega/2}\, b_-(\Omega) \right)\,,
\end{align}
while for the positive P-energy wavefunction we have
\begin{align}\label{fieldx5ap}
\la t|\psi\ra_+ & =\frac{1}{2 \pi} \int^{\infty}_{0}\, {d\Omega}\, \left( t^{-i\Omega}\, \Gamma(i\Omega)\, e^{-\pi\Omega/2} \la \Omega|\psi\rangle + t^{i\Omega}\, \Gamma(-i\Omega)\, e^{\pi\Omega/2} \la -\Omega|\psi\rangle \right)\,,
\end{align}
showing how a purely positive P-frequency field is made of {\it positive} and {\it negative} frequency R-modes.

\section{The thermal distribution}

Let us look back at the action of $ax+b$ transformations on the field \eqref{geoax}
\be\label{geoax}
T(\alpha)\, \psi(t) =  \psi(t + \alpha) \,,\quad  D(\lambda)\,  \psi(t) = \psi(e^{\lambda} t)\,.
\ee
We ask if it is possible to define a new time coordinate on which $D(\lambda)$ acts as a {\it translation}. It is immediate to see that with a change of variable $t=e^{\tau}$ we have 
\be
D(\lambda)\,  \psi(\tau) = \psi(\tau + \lambda)
\ee
while the action of finite translations on this new variable is now slightly involved
\be
T(\alpha)\, \psi(\tau) = \psi(\log(e^\tau+\alpha))\,.
\ee
From these actions we can derive the expression of the generators $P$ and $R$ in terms of the $\tau$-variable
\be
R = - i \frac{d}{d\tau}\,,
\ee
and
\be
P= - i e^{-\tau} \frac{d}{d\tau}\,.
\ee
Some comments are in order at this point: i) the functions of the new variable $\tau$ still carry a legitimate representation of the $ax+b$ group; ii) the change of variables restricts such representation to the positive coordinate axis $t\in \mathbb{R}^+$; iii)  in order to have the dimensions right one should introduce a parameter with dimension of inverse time, $a = [time]^{-1}$, so that $t = \frac{1}{a}\, e^{a\, \tau}$; we set $a=1$ for simplicity but this parameter will be very important at the end. Plane waves oscillating with R-frequency are given by
\be
\la \tau | \Omega\ra_R = \frac{1}{\sqrt{2\pi}}\, e^{i\Omega \tau} = \frac{1}{\sqrt{2\pi}}\, t^{-i\Omega} = {}_+\la t | \Omega\ra_R\,,\,\,\,\, t\in \mathbb{R}^+\,. 
\ee
where we introduced the R-eigenstates $| \Omega\ra_R$. Indeed we have
\be
\la \tau | R | \Omega\ra_R = - i \frac{d}{d\tau} \frac{1}{\sqrt{2\pi}}\, e^{i\Omega \tau} = \Omega\, \la \tau | \Omega\ra_R 
\ee
and
\be
\la \tau | P | \Omega\ra_R = - i e^{-\tau} \frac{d}{d\tau} \frac{1}{\sqrt{2\pi}}\, e^{i\Omega \tau} = \Omega\, \la \tau | \Omega + i \ra_R 
\ee
from which we read the following action of the $ax+b$ generators on the the $| \Omega\ra_R$ states
\be
R\, | \Omega\ra_{R} =  \Omega\, |\Omega\ra_{R} \,,\qquad  P\, | \Omega\ra_{R} = \Omega\, | \Omega + i \ra_{R}\,. 
\ee
This is an irreducible representation of the algebra \eqref{axbalg} on the Hilbert space $\mathcal{H}_{\Omega}=\mathcal{L}^2(\mathbb{R}, d\Omega)$, with $\Omega$ running {\it over the whole real line}, while the representation in terms of the $\tau$ variable is an irreducible representation of the $ax+b$ group on the positive time axis $t \in \mathbb{R}^+$. The representations given by the kets $| \Omega\ra_{R}$, as the $|\Omega\ra$ one, belongs to the class of representations of the $ax+b$ Lie algebra which can be constructed using the so-called Jordan-Schwinger maps from the generators of the Heisenberg algebra \cite{Agostini:2005mf}. 

Let us notice that the functions $t^{i\Omega}$ form a complete set on the positive half line, this can be easily seen by considering the fact that the plane waves $e^{i \Omega \tau}$  provide a complete set of functions for $\tau \in \mathbb{R}$ and thus for $t\in\mathbb{R}^+$. Let us take the restriction of the field to the positive half line and expand it in terms of the $t^{i\omega}$ functions in analogy with \eqref{psikx}
\be\label{ppto}
\psi_{+}(t)\equiv \theta(t)\, \psi(t) = \frac{1}{\sqrt{2 \pi}} \int^{\infty}_{0}\, \frac{d\Omega}{\Omega}\left( t^{i\Omega}\,  c_{+}(\Omega) +  t^{-i\Omega}\,  c_{+}^*(\Omega)\right)\,.
\ee
The space of positive $R$-frequency functions on $\mathbb{R}^+$ will define the (one) ``R-particle" Hilbert space. In the standard Fock space construction $ c_{+}(\Omega)$ and  $c_{+}^*(\Omega)$ become annihilation and creation operators, respectively, and the associated ``R-vacuum" state is defined by
\be
c_{+}(\Omega)  |0\ra_R = 0\,.
\ee
The key question we would like to ask is if such vacuum state is the same state as the P-vacuum $|0\ra_P$. To answer such question let us go back to the expansion \eqref{fieldx2a} of the field in terms of the irreducible representations of the $ax+b$ group diagonal in $P$
\be
\psi(t) = \frac{1}{\sqrt{2 \pi}} \int^{\infty}_{0}\, \frac{d\omega}{\omega}\left( e^{i\omega t}\, {}_+\la \omega | \psi\ra + e^{-i\omega t}\, {}_+\langle \omega| \psi\rangle^* \right)\,,
\ee
and substitute the inverse Mellin transform
\be
 |\omega\rangle_{+} = \int^{\infty}_{-\infty}\, d\Omega\, \langle\Omega |\omega \rangle_+\, |\Omega\rangle = \frac{1}{\sqrt{2 \pi}} \int^{\infty}_{-\infty}\, d\Omega\, \omega^{-i\Omega}\, |\Omega\rangle \,, 
\ee
integrating with respect to $\omega$ we get
\begin{align}
\psi(t) & =\frac{1}{2 \pi} \int^{\infty}_{0}\, {d\Omega}\, t^{-i\Omega}\Gamma(i\Omega)\left( e^{-\pi\Omega/2} \la \Omega|\psi\rangle+ e^{\pi\Omega/2} \la -\Omega|\psi\rangle^* \right) \nonumber\\
&+\frac{1}{2 \pi} \int^{\infty}_{0}\, {d\Omega}\, t^{i\Omega}\Gamma(-i\Omega)\left( e^{-\pi\Omega/2} \la \Omega|\psi\rangle^*+ e^{\pi\Omega/2} \la -\Omega|\psi\rangle \right)\nonumber\,.
\end{align}
From this expression, restricting to positive $t$ and comparing with the $\Omega$ expansion \eqref{ppto}, we obtain
\be
c_+(\Omega) = \frac{\Omega}{\sqrt{2 \pi}}\, \Gamma(-i\Omega)\left( e^{-\pi\Omega/2}\, b^{*}(\Omega)+ e^{\pi\Omega/2}\, b(-\Omega) \right)
\ee
where $b(\Omega) = \la \Omega|\psi\rangle$. It is evident from this relation written in terms of operators on Fock space 
\be\label{bogol}
c_+(\Omega) = \frac{\Omega}{\sqrt{2 \pi}}\, \Gamma(-i\Omega)\left( e^{-\pi\Omega/2}\, b^{\dagger}_+(\Omega)+ e^{\pi\Omega/2}\, b(-\Omega) \right)
\ee
the P-vacuum {\it does not} coincide with the R-vacuum
\be
c_+(\Omega) |0\ra_P \neq 0\,.
\ee
We can actually do more, using \eqref{bogol} we can evaluate the expectation value of the R-particle number operator in the P-vacuum. Taking into account the equality $|\Gamma(i\omega)|^2 = \frac{\pi}{\omega \sinh(\pi \omega)}$, we obtain
\be
{}_P\la 0| c^{\dagger}_+(\omega) c_+(\omega') |0\ra_{P} = \frac{\omega\, \delta(\omega-\omega')}{e^{2\pi\omega}-1}\,,
\ee
showing that the P-vacuum contains {\it a thermal distribution of R-particles} at temperature $1/2 \pi$. As pointed out earlier on in the definition of the time variable $\tau$ in terms of the ``P-time" $t$ we have implicitly introduced a constant $a$ with dimensions of inverse time which for convenience we set to unity in the formulae derived so far. It can be easily checked that writing down explicitly the constant $a$ in the steps above one obtains a thermal distribution exactly at the Unruh temperature $T=a/2 \pi$.

\section{Summary}
We described what in our knowledge is the simplest system in which the freedom in the choice of the generator of time evolution is connected to the appearance of a thermal distribution of excitations similar to the one observed by accelerated observers in the vacuum state of a quantum field in Minkowski space. Thermal effects of this type play a crucial role in semiclassical gravity and are at the basis of outstanding open questions like the nature of black hole entropy and the fate of unitarity in the quantum evolution of black holes \cite{Wald:1999vt,Carlip:2008rk,Mathur:2008wi}. We hope that our derivation can contribute to a deeper understanding of these phenomena providing a minimal setting in which they can be described using only group theoretic inputs concerning the symmetries of the system and their role as quantum observables. Finally it should be pointed out that the $ax+b$ grou plays a key role in a variety of contexts from non-commutative geometry to affine quantization and quantum cosmology \cite{Gayral:2007hk,Klauder:1999ba,Bergeron:2013ika}, it would be interesting to explore the possible relevance of the effect we discussed in this contribution in such disparate fields.


\end{document}